\documentclass[sigconf,screen]{acmart}
\usepackage[ruled,linesnumbered]{algorithm2e}
\usepackage{amsmath}
\usepackage{bbm}

\newcommand{\argmax}[1]{\underset{#1}{\operatorname{arg}\,\operatorname{max}}\;}
\def\BibTeX{{\rm B\kern-.05em{\sc i\kern-.025em b}\kern-.08emT\kern-.1667em\lower.7ex\hbox{E}\kern-.125emX}}
    
\copyrightyear{2019}

\begin{document}

\title[Joint Optimization Framework for Search Experiences]{Revenue, Relevance, Arbitrage and More: Joint Optimization Framework for Search Experiences in Two-Sided Marketplaces}

\author{Andrew Stanton, Akhila Ananthram, Congzhe Su, Liangjie Hong}
\affiliation{%
  \institution{Etsy, Inc}
  \city{New York}
  \country{U.S.A}}
\email{ {astanton, aananthram, csu, lhong}  @ etsy.com}

\begin{abstract}
Two-sided marketplaces such as eBay, Etsy and Taobao have two distinct groups of customers: buyers who use the platform to seek the most relevant and interesting item to purchase and sellers who view the same platform as a tool to reach out to their audience and grow their business. Additionally, platforms have their own objectives ranging from growing both buyer and seller user bases to revenue maximization. It is not difficult to see that it would be challenging to obtain a globally favorable outcome for all parties. Taking the search experience as an example, any interventions are likely to impact either buyers or sellers unfairly to course correct for a greater perceived need. In this paper, we address how a company-aligned search experience can be provided with competing business metrics that E-commerce companies typically tackle. As far as we know, this is a pioneering work to consider multiple different aspects of business indicators in two-sided marketplaces to optimize a search experience. We demonstrate that many problems are difficult or impossible to decompose down to credit assigned scores on individual documents, rendering traditional methods inadequate. Instead, we express market-level metrics as constraints and discuss to what degree multiple potentially conflicting metrics can be tuned to business needs. We further explore the use of policy learners in the form of Evolutionary Strategies to jointly optimize both group-level and market-level metrics simultaneously, side-stepping traditional cascading methods and manual interventions. We empirically evaluate the effectiveness of the proposed method on Etsy data and demonstrate its potential with insights.
\end{abstract}

%
%


%

%
%
\maketitle

\section{Introduction}\label{sec:introduction}

As online shopping becomes a dominant avenue for global buyers, E-commerce companies strive to meet a wide range of often conflicting goals when showing products for their communities. While optimizing for buyer \textit{Conversion Rate}, and therefore higher general revenue or Gross-Merchandise-Value ({\tt GMV}), is commonly considered the top objective, it is usually far from driving a healthy and growing business. In fact, many E-commerce companies, especially those with two-sided marketplaces, face a number of challenges due to over-shackling to {\tt GMV} optimization alone. Without appropriate tempering, such a platform is usually unable to satisfy the short term needs of buyers and sellers as well as the long term needs of the business.

A typical two-sided marketplace such as eBay, Etsy and Taobao has two distinct groups of customers where buyers use the platform to seek the most relevant and interesting item to purchase and sellers view the same platform as a tool to reach out to their audience and grow their business. On top of that, the platform normally would have its own objectives ranging from growing both buyer and seller user bases to {\tt GMV} maximization. It is not difficult to see that it would be challenging to obtain a globally favorable outcome for all parties.

Take showing relevant products to a buyer through the search experience as an example. For a particular purchase intent, or sometimes with a specific item in mind, a buyer would likely discover a spectrum of product listings from multiple sellers in a typical two-sided marketplace. Some seem to be more relevant than others and some even might look the same. A buyer has to decide among these items with a positive experience such that he/she would return to the marketplace next time. For sellers, however, they view search result pages as prominent real estate to gain customers' attention and therefore potentially increase their market share. Maximizing their success in search benefits both their brand and take home pay, regardless of buyers' overall experience on the site or ramifications to other sellers.  On the mission of growing a marketplace, the platform generally needs to step in and sometimes artificially advantage under represented segments of sellers to give them more exposure, creating a reasonably fair competition. However, it might be equally risky to put established sellers in disadvantaged situations, who originally rank well on their own merit, and provide a sub-standard experience to buyers as potentially fewer relevant and lower quality goods are exposed higher through search results.  

This scenario exemplifies a microcosm of the marketplace: any interventions are likely to impact either buyers or sellers unfairly to course correct for a greater perceived need.  On another hand, a two-sided platform also needs to be cautious about the situation where the rich get richer and the poor get poorer, namely the \textit{Matthew Effect}, a factor of constant battle in most marketplaces~\cite{berbeglia2017taming}.  While sellers often perform well due to better commercial strategies and product fit than their peers, a healthy marketplace cannot be solely dominated by a small segment of power sellers and continue to grow. To make things even more complicated, platforms, often operating as modern corporations, subsequently attempt to compensate for these inefficiencies with organizations and teams devoted to their respective customer: for example, buyer, seller, and core market. Hence, as each team attempts to solve a particular problem set, competing needs are demanded of the search experience with each team expecting tuning for their particular business focus. As often these asks are ill-defined and heuristically measured, a grand challenge for building a search ranking algorithm to satisfy all fronts is presented.

In this paper, we address how a company-aligned search experience can be provided with competing business metrics that E-commerce companies typically tackle. As far as we know, this is a pioneering work to consider multiple different aspects of business metrics in two-sided marketplaces to optimize a search experience. We demonstrate that many problems are difficult or impossible to decompose down to credit assigned scores on individual documents, rendering standard point-wise approaches to multi-objective~\cite{svore2011learning} or standard diversity-based\cite{santos2015search} learning to rank algorithms inadequate. Instead, we express market-level metrics as constraints and discuss to what degree multiple potentially conflicting objectives can be tuned to business needs. In addition, we propose a policy learner in the form of \textit{Evolutionary Strategies} to jointly optimize both group-level and market-level metrics simultaneously, side-stepping traditional cascading methods and manual interventions.

The paper is organized as follows. In $\S$\ref{sec:related_work}, we discuss related work in several different directions. We then formulate and define a wide range of metrics including relevancy metrics, diversity metrics and a number of newly proposed market-level metrics relevant to E-commerce interests in Section $\S$\ref{sec:market-metrics}.  We follow up with a proposed set of policies to optimize the above metrics in $\S$\ref{sec:algos}.
Finally in $\S$\ref{sec:experiments}, we empirically evaluate the effectiveness of proposed method on Etsy search logs data, showing how different weightings influence the ultimately delivered rankings. 

\section{Related Work}\label{sec:related_work}

There are number of different facets of work that need to be considered when ranking across a variety of soft constraints.

\subsection{Diversity}

Diversity in Learning to Rank has long and storied past as it relates to Web Search.  The simplest solution typically falls under a heuristic based approach. Carbonell and Goldstein\cite{carbonell1998use} formulated the problem as a selection: Documents are chosen greedily based on a linear combination of query-document relevance and maximal margin relevance (MMR), with each step picking the document which the highest combined score.  After selection, the MMR scores are updated to reflect the newly picked document.  MMR is rooted in the idea of \textit{novelty}, the idea that maximizing the differences in similarity between the set of already selected documents and the remaining set results in a more diverse outcome.  Subsequent work attempts to define better heuristics for selection \cite{santos2015search}.  Dang and Croft propose PM-2 \cite{dang2012diversity}, a diversification method based on proportionality; they argue diversification should be biased toward the overall subtopic proportionality of the entire query-set rather than attempting to balance it uniformly.  xQuAD \cite{santos2010explicit} attempts to understand diversity as a combination of an originating query and derived sub-queries.

In the learning space, the closest related work comes from Xia et al. who describe PAMM \cite{xia2015learning}, a method for optimizing diversity and relevancy via a Perceptron.  Novel to the paper is the idea of direct optimization of the evaluation metrics rather than utilizing heuristics or optimizing surrogate functions.  PAMM works generally by sampling both positively and negatively ranked lists and attempts to maximize the margin between them.

Crucially, all of the above attempt to solve the challenge of query ambiguity; the idea that redundancy is undesired due to lack of strong conviction of the topicality of a given query.  This positions the frameworks to satisfy the needs of the searcher, with the underlying objective of improving the browser's experience. While two-sided marketplaces face similar challenges to query ambiguity, they also need to balance the needs of the seller: diversity for the sake of mitigating bias.  To our knowledge, none of the proposed work attempt to optimize for market-level diversity metrics.

\subsection{Policy Learning}

Policy optimization in LTR space has come in a few flavors over the years, with most of its history focused in online LTR.  Radlinski et al. \cite{radlinski2008learning} first discussed diversity-based online optimization in the form of using multi-armed bandits to minimize page abandonment. They condition the expected reward on previous documents selected, considering each remaining document an "arm", allowing for suitable exploration/exploitation trade-off against the expected reward.  More recent work proposes modeling user behavior as an MDP\cite{hu2018reinforcement} with the goal of learning how browser sessions can be utilized in re-ranking.  They proceed to describe a policy gradient method to learn optimal ranking policies given the learned SS-MDP.

Most applicable to our proposed policy is utilizing black box optimization in learning to rank.  Salimans et al. recently showed that Evolutionary Strategies \cite{salimans2017evolution} were well suited for learning reinforcement problems, applying a variation of ES known as Natural Evolutionary Strategies\cite{wierstra2008natural} to Atari game learning.  They proposed a scalable algorithm and demonstrated the optimizer's tolerance to stochastic environments.   \cite{chrabaszcz2018back} showed competitive results to Salimans via \textit{Canonical ES} - a simpler version of the ($\lambda$, 1) variants of Evolutionary Strategies. Concurrently, Ibrahim et al. applied perhaps the simplest type of  (1+1)-Evolutionary Strategies to learn policies on linear models to directly optimize the average nDCG across all query sets\cite{ibrahim2018evolutionary}.  
%

\subsection{Popularity Bias and the Matthew Effect}

Measuring market level performance within search is fairly under researched in the space of E-Commerce.  Perhaps closest to our particular use case is from Hentenryck et al. \cite{van2016aligning} who analyzed the impacts of social influence on a trial-offer market, showing that ranking on conditional purchase rate lead to natural monopolies by the highest quality products.  Follow up work attempts to compensate for the Matthew Effect \cite{berbeglia2017taming} by intervening with a stochastic policy to randomize products of similar quality.  They show how segmenting products into different "worlds" and conditioning popularity on which world a user observes results in a stable market where products of similar quality obtain equivalent market share.

\section{Metrics For Optimization}\label{sec:market-metrics}
In this section, we outline metrics for optimization from a typical two-sided marketplace. We review classic relevancy metrics in $\S$\ref{sec:metrics_relevancy} as well as diversity metrics in $\S$\ref{sec:metrics_diversity}, serving the foundation of metrics for modern search ranking. We introduce a new class of market-level metrics in $\S$\ref{sec:metrics_market}. We list all notations used through out the paper in Table \ref{table:notations}.
\begin{table}[h]
\caption{Notation}
\begin{tabular}{ c||l}
 \textbf{Notation} & \textbf{Description} \\ \hline \hline
 $ Q = \{q_1,...,q_n\} $ & Unique query set \\ 
 $ N_i $ & Number of documents in query set $q_i$ \\ 
 $ D = \{d_1,...,d_j\} $ & Document set \\ 
 $ Feats(d_i) $ & Feature vector for document $d_i$ \\
 $ Y \in \{1,2,3,4,5\} $ & Relevance grades  \\
 $y_{i,j} \in Y$ & Relevance score for $q_i$ and $d_j$ \\
 $ \pi $ & Ranking policy \\
 $ R(\pi, q_i) = <d_1, \ldots)>$ & Ranked documents for query $q_i$ \\
 $ V(\pi, q_i, d_j)$ & Evaluator for query-doc set \\
 $ S(\pi, Q, D) \in  [0,1] $ & Market-level objective function \\
 \hline
\end{tabular}
\label{table:notations}
\end{table}
\subsection{Relevancy Metrics}\label{sec:metrics_relevancy}
First and foremost is the concept of relevancy, rooted originally in the well-known \textit{Probability Ranking Principle} ({\tt PRP})  framework~\cite{robertson1977probability,cooper1971prp} which states that documents should be ordered independently in decreasing presentation of relevance.  That is, the most relevant document for a query should be placed first. To measure that principal, industry has standardized around two core metrics for evaluating the efficacy of their ranking systems: {\tt NDCG}~\cite{jarvelin2000ir} and {\tt ERR}~\cite{chapelle2009expected}.  We now give a brief overview of their formulation.

\medskip

{\bfseries Normalized Discounted Cumulative Gain (NDCG)}:  {\tt NDCG} is an ordered relevance metric measuring the agreement between a goldset list of documents and the permutation return by the ranking policy.  It is typically evaluated to some position $K$, indicating only the first $K$ documents should be considered for evaluation.  Usually values are small, emphasizing the importance of getting the first few documents correct.  Given a query $q_i$:
\begin{align} \label{NDCG}
\begin{split} 
   \mathrm{DCG@K} &= \sum_{j=1}^{K} \frac{2^{y_{i,j}} - 1 }{\log_2(i + 1)} \\
   \mathrm{NDCG@K} &= \frac{\mathrm{DCG@K}}{\mathrm{IDCG@K}} \in [0,1]
\end{split}
\end{align}Where {\tt IDCG} is the best possible {\tt DCG} score for the given query-doc set.  We typically evaluate the average {\tt NDCG@K} across all queries.
  
\medskip

{\bfseries Expected Reciprocal Rank (ERR)}: {\tt ERR}~\cite{chapelle2009expected} is proposed as an adjustment to {\tt NDCG}, attempting to factor in a prior to how users actually consider documents for engagement.  While graded labels are still assigned to documents independently, {\tt ERR} is grounded on the idea of the cascade user model~\cite{craswell2008experimental}: that previous evaluations of documents influences the likelihood that a buyer will continue browsing.  While a full discussion on {\tt ERR} is out of the scope of this paper, we provide the following formulation used in evaluation.  Given some mapping $R$ of relevance grades to probability of relevance,  we can define {\tt ERR} as:
\begin{align*}
    R(g) &= \frac{2^{g} - 1}{2^{\max (Y)}} \\
    p_0 &= 1, p_j = p_{j-1}(1 - R(y_{i,j})) \\
    \mathrm{ERR}_0 &= 0, \mathrm{ERR}_j = \mathrm{ERR}_{j-1} + p_{j-1}\frac{R(y_{i,j})}{j}
\end{align*}
\subsection{Diversity Metrics}\label{sec:metrics_diversity}
Evaluation of sub-topic diversity is a rich field with many contributions ~\cite{zhai2015beyond,agrawal2009diversifying,chapelle2011intent,clarke2008novelty}.  Generally, the metrics revolve around the idea that there exist ambiguity in the intent behind a query.  For example, while there is likely a strong relationship between the query ``lace bridal veil'' and the {\tt /clothing/wedding/accessory/veils} taxonomy, for other queries there is less implicit understanding.  On Etsy, we serve a large number of inspirational (cheerful, happy, beautiful), stylistic (geometric, upcycled, animal print), and occasion (gifts for him, bridesmaid presents, stocking stuffers) queries which have high taxonomic (interchangeably used with topicality) ambiguity.  In cases where there is low certainty of strong topicality, ranking benefits from increasing coverage of different sub topics early on in the presented result set.  Indeed, empirical results have shown increasing diversity improves user engagement metrics~\cite{raman2012learning}.

\medskip

{\bfseries ERR-IA}: {\tt ERR-IA}~\cite{chapelle2009expected} is an extension to {\tt ERR} that incorporates the notion of diversity. Let $\Pr( t \mid q )$ be the probability of a topic, t, for a given query, q. We can define {\tt ERR-IA} as:
\[
R(y_i, t) =  \begin{cases} 
      R(y_i) & i\in t \\
      0 & else
   \end{cases}
\]
\[
\mathrm{ERRIA@K} = \sum_{j=1}^{K} \frac{1}{j} \sum_{t} \Pr(t \mid q)\prod_{i=1}^{j-1}(1 - R(y_i, t)) R(y_j, t)
\]
Or, using the {\tt ERR} formula from before, we can define it as:
\begin{equation} \label{err-ia}
\begin{split}
\mathrm{scores} = [y_i * \mathbbm{1}{[i \in cat_t]}, \forall i \in [1,K]]\\
\mathrm{ERRIA@K} = \sum_{t} \Pr(t \mid q) * \mathrm{ERR@K}(\mathrm{scores})
\end{split}
\end{equation}

\subsection{Market-Level Metrics}\label{sec:metrics_market}
Two sided marketplaces, unlike traditional web search, suffers from multiple challenges typically framed in the form of \textit{inequality}; the realization that there exists some skew in the marketplace we wish to correct.  In this sub section we introduce a wide range of different types of market corrections and present potential metrics for optimization. Before we introduce the first proposed metric, we discuss the notion of query-set dependence below.

\medskip

{\bfseries Query-Set Dependence}: At first glance, many metrics seem like they could be modeled as group-wise diversity problems. Let us consider the case of balancing between highly successful power sellers and new sellers on the site.  While majority of the sales will come from a fraction of the entire seller base, an over saturation of a small proportion of sellers in search has the potential to squeeze out new shops on the site.  As there is strong evidence that faster time to first sale increases a seller's \textit{Life Time Value} ({\tt LTV}), we would ideally like to improve their exposure.  Applying a diversity constraint between power sellers and new sellers seems reasonable: for each query set, increase coverage between the two distributions in the top $K$ spots.  However, this ignores a fundamental problem: different queries have different amounts of traffic associated with them, preventing us from properly balancing the market.  Much like the {\tt PRP} models before, diversity metrics assume query-level independence: we only consider how the impact of diversification adjusts the metrics within each query set, not its contribution to the overall market place. Another simple example can illustrate the differences.  If we have two ranked sets of items, where $N$ indicates a new seller and $P$ indicates a power seller, for rankings $\{P,P,N,N\}$ and $\{N,N,P,P\}$, the average diversity scores for both queries will be low.  However, from an inequality perspective, both power sellers and new sellers have equal representation in top spots, satisfying the market level needs. Overall, we can write our general market indicator function with only a slight abuse of notation as:
\begin{equation} \label{EQ:Indicator}
I(S | Q_{\pi}), S \in [0,1]    
\end{equation}where $Q_{\pi}$ is the rankings of all documents across all queries by policy $\pi$.

\subsubsection{Weighted Importance Ranking}
This leads us to our first proposed metric: Weighted Importance Ranking.  Under the assumption that we want to balance some scoring function $S_{sub}$ according to some importance weighting, such as traffic volume, we can describe our score function S as:
\begin{equation} \label{ImpWeighting}
S(\pi) = \frac 
{\sum_{i=1}^{N} W_i * S_{sub}(\pi,q_i,d_i)}
{\sum_{i=1}^{N} W_i}
\end{equation}This provides a useful feedback mechanism to the policy learner: much like the intuition that rankers optimizing for {\tt NDCG} or {\tt ERR} should focus their energy on improving the rank of documents higher in the page,  providing importance feedback provides guidance to the policy on which queries it should focus time on optimizing.

\subsubsection{Outlier Skew}
When dealing with implicit feedback data, chronic cold starts, or otherwise uncertainty in the relevance set, optimizing for the expected {\tt NDCG} across all queries can lead to an over-sensitivity to outliers: query sets that are either trivial or impossible to successfully rank given the features.

To address the influence, we propose a simple change to the optimization function to maximize the scores at given percentiles instead of the mean.  Given the set of scores, $M$, a set of percentiles to evaluate (e.g. 25th, 75th, etc.), $\mathrm{Percentiles}$:
\begin{equation} \label{Histogram}
\begin{split}
M &= \{V(\pi, q_1, d_q),...,V(\pi,q_N,d_N)\} \\
 S(\pi) &= \frac 
{\sum_{p \in \mathrm{Percentiles}}M_(p)}
{|\mathrm{Percentiles}|}
\end{split}
\end{equation}Empirically, we have found that optimizing quantiles can provide a smoother distribution. 

\subsubsection{Incentives}
Many cases where business wishes to correct market bias can boil down to minimizing arbitrage.  For example, it might be observed that Buyers find items with low list prices attractive and yet are surprised when confronted with costly shipping.  Sellers will often discover these user behaviors can yield a higher collection of clicks, a standard ranking signal, and will list items with artificially low prices to improve their ranking in Search.

Incorporating incentives in a principled way is difficult. As a marketplace which caters to many different price points, many of which are materially cheap to produce, this rules out simple heuristics or other hard rules.  Analysis might come up with some rule, say listing prices lower than the median value of a product should be down weighted in search (also known as \textit{burying}), but that goes too far: we now penalize sellers producing goods more efficiently.

Rather than exert a hard penalty as a group-wise metric, we instead propose a simple maximization approach across the top ranked documents for all queries. Given some user-provided behavior function, $B$, indicating that a document for a given query exhibits a quality we wish to incentivize, a ranking of documents for query $q_i$, and the number of positions, $K$, to consider:

\begin{equation} \label{Incentive}
\begin{split}
pos &\in \{1,...,N_i\} \\
B(q_i, d_j) &= {0,1} \\
S(\pi) &= \frac{\sum_{i=1}^{|Q|} \sum_{p=0}^{K} B(q_i, R(\pi, q_i)@p)} {K|Q|}
\end{split} 
\end{equation}Combined with Weighted Importance Ranking in Equation \ref{ImpWeighting}, we can influence desired behaviors conditioned on relevance and group-level diversity.

\subsubsection{Inequality}
\begin{figure}
    \centering
    \includegraphics[scale=0.5]{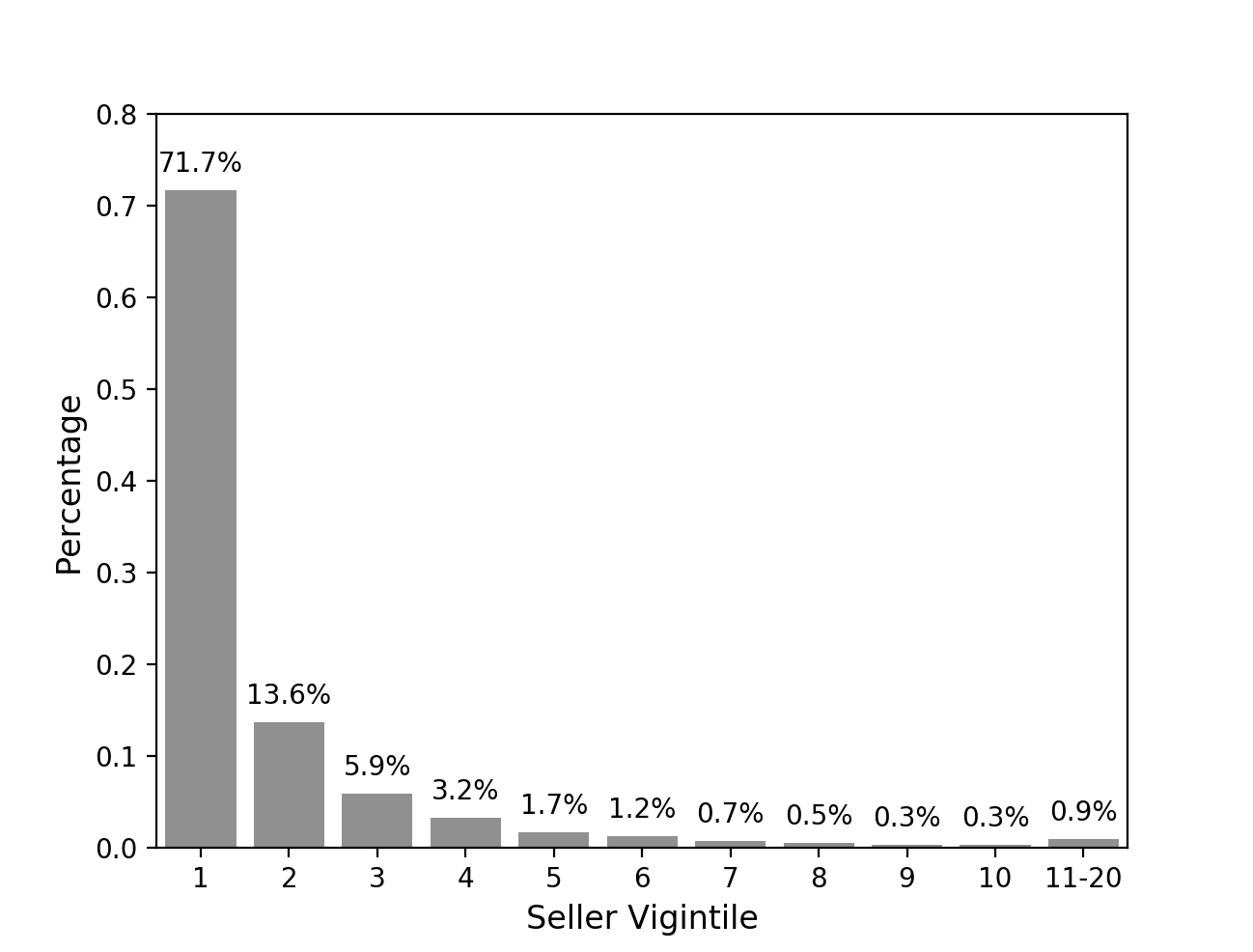}
    \caption{Seller Distribution}
    \label{fig:seller-distribution}
\end{figure}
The final class of market level metrics fall under the guise of \textit{inequality}: some imbalance in wealth distribution across tiers of sellers that we wish to correct.  Without interventions, marketplaces often fall under the phenomena of the Matthew Effect; wealth accumulates in a small segment of the population, squeezing out other Sellers.  Etsy is no different: Figure \ref{fig:seller-distribution} shows how visibility accumulates in the wealthy few: the top few\% of sellers account for the bulk of sales.  Indeed, it can be shown that models that maximizing conditional purchase rate will result in natural monopolies~\cite{van2016aligning} due to social signals present in the marketplace (reviews, best sellers, number of sales, etc.).  While beneficial to maximizing sales, it carries inherent business risk: any loss of sellers occupying the monopoly positions for popular queries will have outsize effect on bottom line {\tt KPIs}.

While work has been done examining how different rankings can impact a market's health (e.g. maximizing number of purchases)~\cite{berbeglia2017taming}, there is little discussion on improving proportional representation across tiers of sellers jointly.

Given our desire to influence inequality, we examine two methods for flattening the distribution, depending on the amount of prior information we have.

\medskip

{\bfseries Gini Index}: In the case where we can estimate or measure the wealth distribution apriori, we propose minimizing the Gini Index\cite{gini1912italian}, also known as the Gini Coefficient, across all query sets.  Based on the Lorenz Curve, it estimates the difference between full income equality and the actual observed wealth distribution of different sub populations. Given a cumulative proportion of the population, $X_i$ and a cumulative proportion of wealth $W_i$, ordered such that $w_i / x_i \leq w_{i+1}/ x_{i+1}$ we can define the Gini Index indicator:

\begin{align*}
    \mathrm{Gini} &= 1 - \sum_{i=2}^{|X|} (X_i - X_{i-1})(\mathrm{W}_i + \mathrm{W}_{i-1}) \\
    S(\pi) &= 1 - \mathrm{Gini}
\end{align*}

There are a number of different ways to compute the population: total count of different sub populations across all inventory, such as seller deciles, the subset represented in the train dataset, etc.  Similarly, there are many ways we can measure wealth; for E-commerce businesses, purchase count per query can be a reasonable approximation for seller wealth where, given some function $\mathrm{IsSubPop}$:

\begin{align*}
    w_i &= \sum_{j=1}^{|Q|} \mathrm{IsSubPop}(\pi, q_j, i) \cdot Purchases(q_j)
\end{align*}

In the case where we are more interested in the traffic distribution, W can be set to the amount of query volume. For simplicity, in our experiments, we compute Gini for rank=1. However, computing the Gini Coefficient over multiple rank positions is similarly straight forward if we have access to the observation probability O of a document at position p (such as from a click model):

\begin{equation} \label{MultiGini}
\begin{split}
    O(\mathrm{Pos}) &\in [0,1] \\
    S_{\mathrm{pos}}(\pi) &= O(p) * \mathrm{Gini}
\end{split}
\end{equation}

\medskip

{\bfseries $\mathcal{\chi}^2$ Uniformity}: In cases where P and W are unknown, we can use a simpler method for measuring inequality: maximizing the inverse of the $\chi^2$ fit against a uniform distribution:
\begin{equation} \label{ChiSquared}
\begin{split}
    \mathrm{Count}(\pi,cat) &= {\sum_{p=1}^K \sum_{i=1}^{|Q|} \mathrm{IsCat}(\pi, R(\pi,q_i, p),\mathrm{cat})} \\
    \mathcal{\chi}_{uniform}^2 &= \frac{\sum_{c \in C} (\mathrm{Count}(\pi, c) - \frac {|Q|}{|C|})^2 }{\frac {|Q|}{|C|}} \\
    S(\pi) &= \frac{1}{1 + \mathcal{\chi}_{\mathrm{Uniform}}^2}
\end{split}
\end{equation}

\section{Algorithm Overview} \label{sec:algos}

Learning a ranking policy requires multiple levels of information: individual scores in the case of relevancy, group-level metrics for diversity, and market-level to account for a variety of skews.  Further difficulty arises from the large number of rank orderings: many metrics are neither continuous or differentiable. 

To optimize the policies, we compose a linear combination of all metrics (e.g. NDCG, incentives, gini, etc) into a final \textit{Fitness Function}, which our proposed optimizer tries to maximize. 

\begin{equation}\label{eq:linear-combo}
    F(\pi) = \frac{\sum_{i=1}^{|S|} W_i \cdot S_i(\pi)}{\sum_{i=1}^{|S|} W_i}
\end{equation}

Below we describe a greedy, group-level policy optimizing either a static or stochastic value function followed by a proposed optimizer to learn the policies.

\subsection{Greedy Algorithm}

As has been shown through numerous previous works\cite{santos2015search}, the assumption of independence during prediction is violated when considering group-level diversity.  Consequently, most work on diversity utilize a heuristic, second pass algorithm to select subsequent documents during the ranking process.  However, Zhai et al. showed that a simple greedy algorithm performed well in maximizing MMR\cite{zhai2015beyond}\ref{Greedy-Algo}:

\begin{algorithm}[h]
\SetAlgoNoLine
\textbf{Input: } parameters $\sigma$, value function $v$, documents $D$ \\
\textbf{Output: } $\langle d_1, d_2, ..., d_k \rangle$ \\
\For{i = {1, 2, 3, ..., K}}{
    $d_i = \argmax{d \in D} v(\sigma, d_i, \langle d_1, d_2,..., d_{i-1} \rangle) $ \\
    $ D = D - \{d_i\} $
}
\caption{Greedy Algorithm}
\label{Greedy-Algo}
\end{algorithm}

\subsubsection{Static Value Functions}

While the greedy policy classically focuses on a heuristic document similarity as the selection criteria for the value function, we instead learn a parameterization over the greedy algorithm and utilize a simple average over the features of previously selected documents to represent aggregate state.

We define a static value function as:

\begin{equation} \label{Static-Value-Function}
\begin{split} 
    s_1 = 0, 
    s_i &= \frac{\sum_{j=1}^{i-1} Feats(d_i)}{i - 1} \\
    v(\sigma, d_i, \langle d_1, .., d_{i-1} \rangle) &= \Phi(\sigma, s_{i-1} - Feats(d_i))
\end{split}
\end{equation}

where $\Phi$ is a fully connected neural network.

\subsubsection{Stochastic Value Functions}

In this section we introduce a stochastic value function, SVF for short.  Queries are assumed independent much the same way that the PRP assumed documents can be arranged independently of each other.  However, it's easy to see how this assumption is violated.

Revisiting the previous example of new sellers and power sellers, we present a set of rankings from two policies $\pi_1$ and $\pi_2$:

\begin{align*}
  Q1_{\pi_1} &= \{P, N, P, N\}, 
  Q2_{\pi_1} = \{P, N, P, N\} \\
  Q1_{\pi_2} &= \{N, P, N, P\}, 
  Q2_{\pi_2} = \{N, P, N, P\} 
\end{align*}
  
It is clear that the group level diversity metrics for each policy are optimal given these two seller categories, but also equally clear that the diversity of sellers occupying the first position is poor.  In expectation, one can also see how blending the two policies would result in the highest overall reward:\[\pi = \argmax{p \in \{\pi_1, \pi_2\}} \operatorname{Uniform}(0, 1)) \]

Inspired by the observation that neural networks can be viewed as an exponential set of sub-networks \cite{srivastava2014dropout} and the above observation, we introduce a simple stochastic feature into the network to allow for the blending of learned sub-policies:

\begin{equation} \label{Stochastic-Value-Function}
\begin{split} 
    f &\sim \operatorname{Uniform}(0, 1) \\
    v(\sigma, d_i, \langle d_1, .., d_{i-1} \rangle) &= \Phi(\sigma, (s_{i-1} - Feats(d_i)) \oplus f)
\end{split}
\end{equation}

Rather than break apart $\theta$ into explicit policy sets, we rely on feature masking in \ref{General-ES} to produce thinned, sub-networks for optimization, deferring to the optimizer to learn how best to incorporate the noise.

\subsection{Evolutionary Strategies}
Learning a policy that devolves into various forms of sorting is difficult; given slight variations to the underlying parameters can result in large swings in scores.  To solve this challenge, we reach for recent work using ES from the LTR and Reinforcement Learning space to maximize our desired objectives.  Below we provide a high level description of the $(1 + \lambda)$ variety of ES.

Given a parameter set $\theta$ (henceforth known as the parent), a fitness function $f$, and shaping function $H$, we sample $\lambda$ search gradients from the Normal distribution: $ i \in \{1, 2, .., \lambda\}, \epsilon_i \sim \mathcal{N}(\mu, I) $ where $\mu = 0$ and $I = 1$ are typical parameters for the noise distribution; while both Salimans and Chrabaszcz explored adjusting $I$, neither found it significantly changed the results.  We further augment the algorithm by masking parameters with probability $({p})$, reducing the effective search space per pass~\cite{ibrahim2018evolutionary}. For each search gradient, we compute its fitness with respect to the parent $\mathrm{Fitness}_i = f(\theta_{\mathrm{parent}} + \epsilon_i)$.  We proceed to run all scores through a shaping function which scales each gradient by some rank function to smooth out the impact of outlier fitness scores: $\epsilon'_i = \epsilon_i * H(\mathrm{Fitness}_i, \mathrm{Fitness}_*)$.  We compute our candidate parent as the sum of gradients scaled by $\sigma$ and compare it to the previous parent, replacing the parent if the candidate improves.
\begin{align*}
   P &= \{\theta_{\mathrm{parent}}, \theta_{\mathrm{parent}} + \sigma * \sum_{i=1}^{\lambda} \epsilon'_i\} \\
   \theta &= {\argmax{p \in P} F(p)}
\end{align*}There are a few variations which are commonly used.  The first is always updating $\theta_{\mathrm{parent}} = \theta_{\mathrm{candidate}}$ regardless of improvement of fitness.  The second is with respect to the shaping function: Wierstra et al. explored the impact of fitness shaping functions in Natural Evolutionary Strategies~\cite{wierstra2008natural} and found that so long as they were monotonic with respect to utility rank, they improved the robustness.  The final one is the number of Search gradients used during candidate construction, which also correspond to the theoretical underpinnings: Salimans et al. used all gradients as part of its computation due to assumptions made in NES whereas~\cite{chrabaszcz2018back} implement a canonical variant which only uses the best $\mu$ children.

We consolidate all of these variants into the following generalized $(1+\lambda)-ES$ algorithm in \ref{General-ES}.

\begin{algorithm}[h]
\textbf{Input: } 
$\theta_0$ - parameters,
F - fitness function,  
H - shaping function,
$\left({p}\right)$ - mask probability,
update $\in \{True,False\}$, 
$\lambda \in \mathcal{I}^{+}$, 
$\mu \in \mathcal{I}^{+}$, 
iters $\in \mathcal{I}^{+}$ \\
\For{i = \{1, 2, .., iters\}} {
    \For{c = \{1..$\lambda$\}} {
        $\epsilon_c \sim \mathcal{N}(0,1) \cdot \operatorname{Bern} \left({p}\right) $ \\
        $ s_c = F(\theta_{i-1} + \epsilon_c)$ \\
    }
    $Sort(\epsilon_*, s_*) \text{ in non-increasing order } \{s_1 \geq s_2 \geq .. s_{\lambda}\} $ \\
    $ \theta_{candidate} = \theta_{i-1} + \sum_{j=1}^{\mu} \epsilon_j \cdot H(s_j, \{s_1,s_2,.., s_{\mu}\}) $ \\
    \uIf{\textit{update}}{
        $ \theta_i \leftarrow \theta_{candidate} $ \\
    }\Else{
        $P = \{\theta_{i-1}, \theta_{candidate}\} $ \\
        $\theta_i \leftarrow P_{\argmax{p \in P} F(p)} $ \\
    }
}
\caption{Generalized $ (1-\lambda) $-ES}
\label{General-ES}
\end{algorithm}

We add a few additions on top of the Generalized ES algorithm.  First we observe that the Greedy policy complexity is $\mathcal{O}(N^2)$, leading to severe slowdown on large document sets during training.  To mitigate this, we uniformly sub-sample the document set each pass, for each query.  While left as a hyper parameter, we found setting it to twice the K value used to compute NDCG sufficient for convergence. Secondly, we add batching instead of optimizing the entire dataset at once. Finally, we utilize the the masking strategy as in ES-Rank\cite{ibrahim2018evolutionary} except that we sample from the Bernoulli distribution with some probability $\left({p}\right)$.  
\section{Experiments}\label{sec:experiments}
In this section, we firstly discuss the data we use in experiments in $\S$\ref{sec:exp_data} and then we discuss our baselines and chosen implementation details in $\S$\ref{sec:exp_algo} and finally we analyze results in $\S$\ref{sec:experimental_analysis}.

\subsection{Data}\label{sec:exp_data}
\begin{figure}
    \centering
    \includegraphics[scale=0.13]{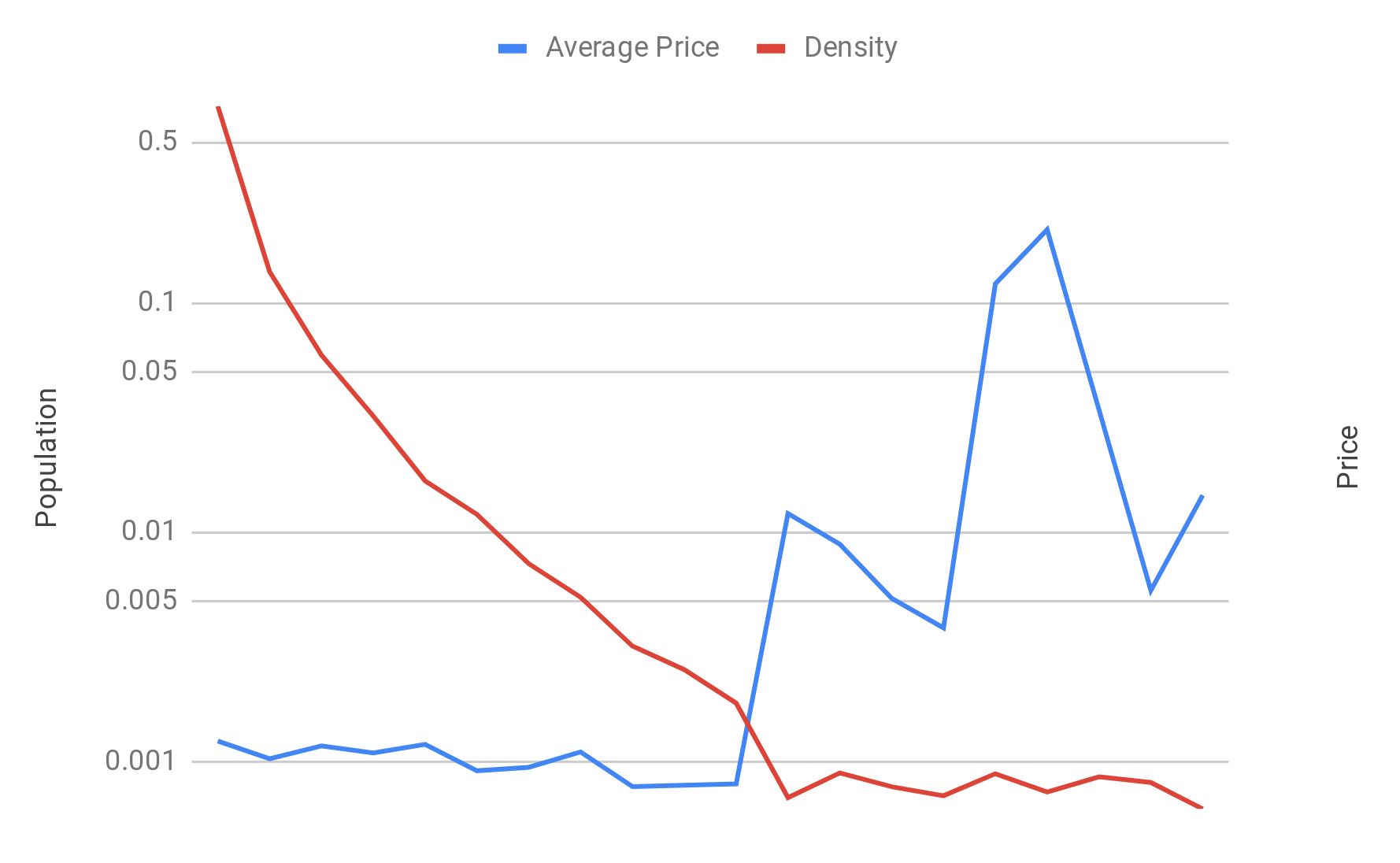}
    \caption{Population vs Price}
    \label{fig:pop-price}
\end{figure}
Our training set is extracted from production search logs, containing the top 5,000 queries observed over an eight day period.  Document features are a combination of query relevancy, historical performance, and taxonomic information. We further augment the dataset with a variety of metadata:

\medskip

\noindent
{\bfseries Population, Wealth, Observation}: Seller population scores are based on {\tt GMV} vigintiles: Etsy, like most e-commerce sites, exhibits a heavy power law distribution with respect to dollar shares.  We compute the observation model by summing the number of purchases over a week at each rank position and dividing by the total number of purchases observed. Wealth is calculated by summing the total number of purchases per query over the same week as the observation model.

\medskip

\noindent
{\bfseries Diversity}: Listings taxonomy is used as the source of diversity in queries.  Overall, it has 175 different categories with a large class imbalance.

\medskip

\begin{figure}
    \centering
    \includegraphics[scale=0.5]{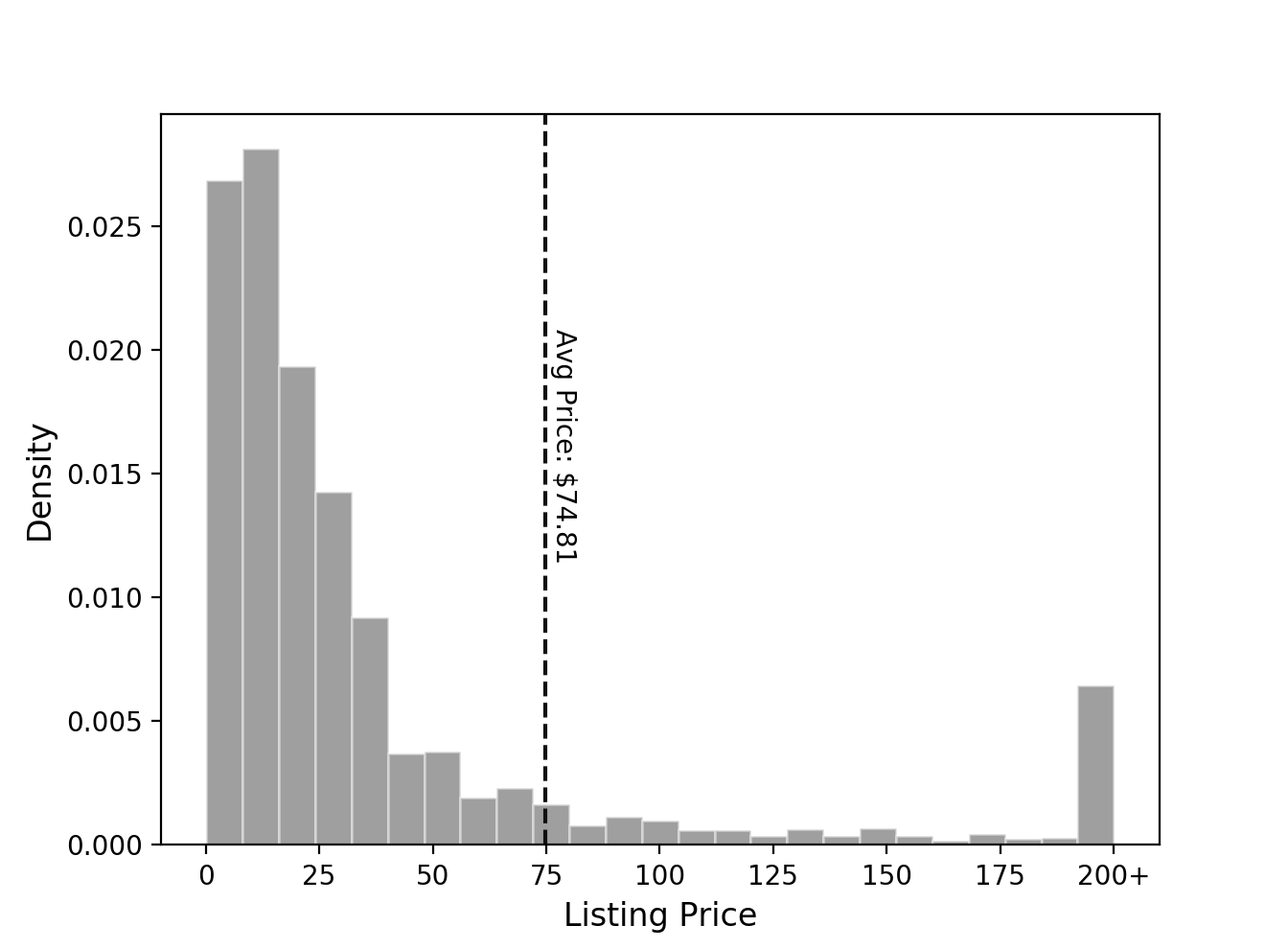}
    \caption{Price Distribution}
    \label{fig:price}
\end{figure}

\noindent
{\bfseries Incentives}:  We binarize our product prices by the mean listing value in our sampled search results.  Price has a sharp skew (\ref{fig:price}) toward lower cost items, often times washing out higher quality, more labor intensive products.  We add an incentive toward premium prices to boost high quality listings closer to the top of the rankings.
\medskip

\noindent
{\bfseries Gini vs Price}: As can be seen in figure (\ref{fig:pop-price}), listing price and population counts are inversely correlated, illustrating a common problem faced by e-commerce: improving the Gini Index naturally results in a reduction in listing price.

\subsection{Algorithms}\label{sec:exp_algo}
\begin{table}[ht]
\scriptsize
\caption{Weight Variants}
 \begin{tabular}{l || c | c c c} 
 Variant & Relevance & Group Diversity & Gini Index & Incentive \\ 
 \hline\hline
 -\{0,0\} & 1.00 & 0.00 & 0.00 & 0.00 \\ 
 \hline
 -\{0.05,0.05\} & 0.85 & 0.05 & 0.05 & 0.05 \\
 \hline
 -\{0.1,0.1\} & 0.70 & 0.10 & 0.10 & 0.10 \\
 \hline
 -\{0.17,0.17\} & 0.49 & 0.17 & 0.17 & 0.17 \\
 \hline
 -\{0.25,0.25\} & 0.25 & 0.25 & 0.25 & 0.25 \\
 \hline
 -\{0.3,0.3\} & 0.10 & 0.30 & 0.30 & 0.30 \\
  \hline
 -\{0.05,0\} & 0.90 & 0.00 & 0.05 & 0.05 \\
  \hline
 -\{0.1,0\} & 0.80 & 0.00 & 0.10 & 0.10 \\
  \hline
 -\{0.25,0\} & 0.50 & 0.00 & 0.25 & 0.25 \\
  \hline
 -\{0.33,0\} & 0.33 & 0.00 & 0.33 & 0.33 \\
  \hline
 -\{0.4,0\} & 0.20 & 0.00 & 0.40 & 0.40 \\
 [1ex] 
\end{tabular}
\label{Tab:weights}
\end{table}

\medskip

\noindent
{\bfseries Baselines}: We compare two different baselines to affirm efficacy of our approach.  We first look at the venerable Maximal Marginal Relevance \cite{carbonell1998use} where the relevance model is learned using LambdaMART \cite{burges2010ranknet} optimized for {\tt NDCG@10}.  Document similarity is based on the Jaccard of each document's taxonomy.  We tune the blend parameter $\lambda$ by maximizing the weighted sum of scores between all indicators. The second model is optimized using the same generalized ES algorithm above, however we replace the greedy policy with a standard pointwise inference policy, sorting documents based on their learned scores.

\medskip

\noindent
{\bfseries Evolutionary Strategies}: In our experiments we compare ES policies trained against a variety of different metrics:
\begin{itemize}
    \item Relevancy scores measured as {\tt NDCG@10} (\ref{NDCG})
    \item Groupwise diversity using {\tt ERR-IA@10} (\ref{err-ia}) across different taxonomic groups 
    \item Market indicators: $\mathrm{Gini}_1$ and $\mathrm{Incentives}_1$ (\ref{Incentive})
\end{itemize}We combine these into our fitness function, $F$, via a weighted linear combination in EQ \ref{eq:linear-combo} with weights described in Table \ref{Tab:weights}. We explore two different policies: a standard point-wise baseline and a greedy policy (\ref{Greedy-Algo}). We use the generalized form of Canonical Evolutionary Strategies, fixing $\lambda=768$ and $\mu=50$ for all tests.  We set \textit{update} to True and fix  $\left({p}\right)$ at 0.05. We compare three different variations of our proposed models: Greedy with a static value function, Greedy with a stochastic value function, and Pointwise with a stochastic value function.  For $\Phi$, we use a small fully connected neural net (20 $\rightarrow$ 20 $\rightarrow$ 1) utilizing the ReLU\cite{jarrett2009best} non-linearity. When utilizing stochastic value functions, we evaluate the test set 5 times with different random seeds to determine expected performance.

\subsection{Experimental Analysis}\label{sec:experimental_analysis}
\begin{figure}
    \centering
     \includegraphics[width=90mm,scale=0.5]{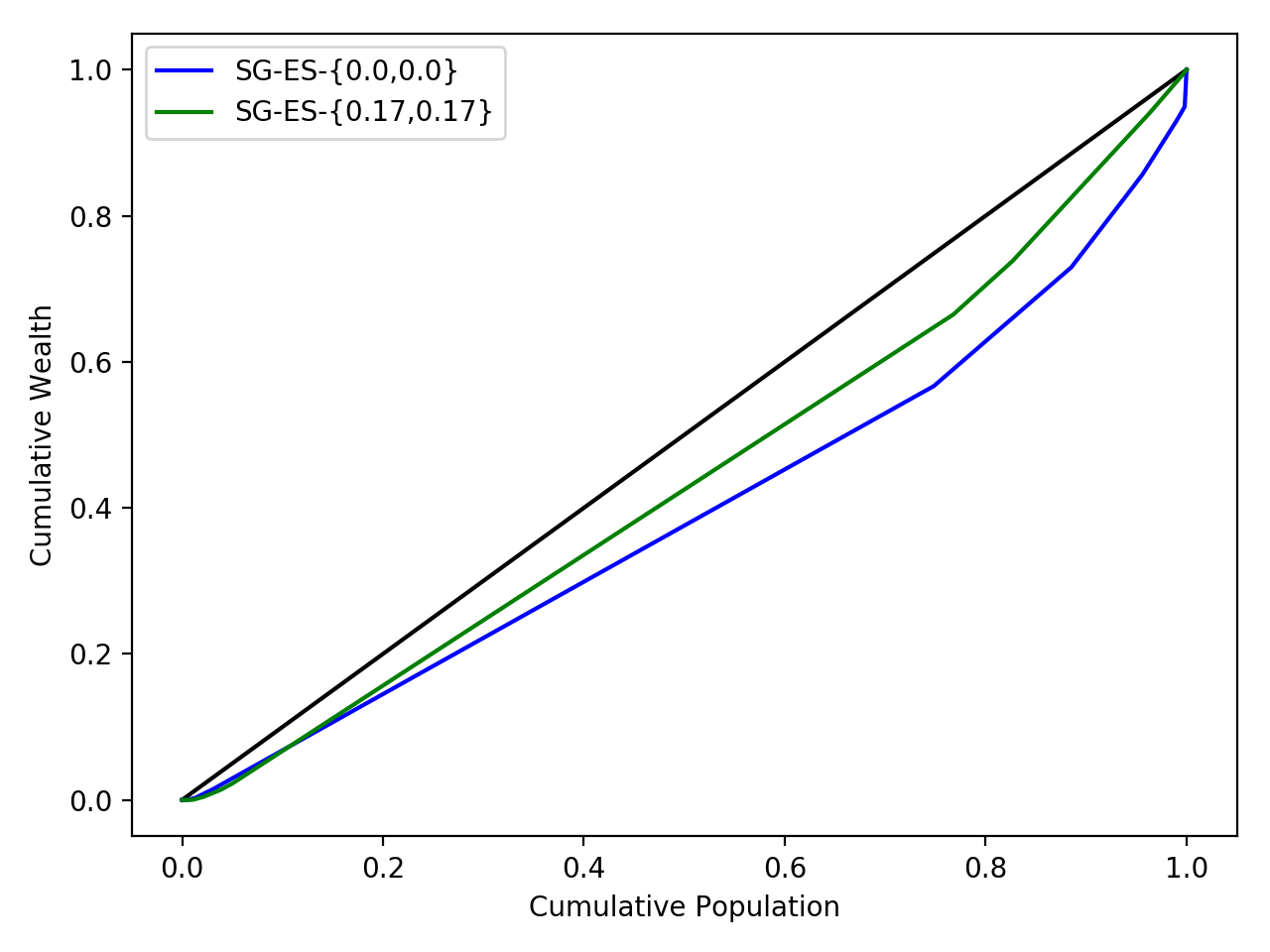}
    \caption{Lorenz Curve (SG-ES)}
    \label{fig:lorenz}
\end{figure}
To verify efficacy of our approach, we examine a few different axes in fitness.  To tease apart how different weights impact the overall model, we first examine how market indicator constraints impact relevancy across the different policies.  We follow it up by examining to what degree relevancy, group-level, and indicators can jointly optimize all metrics simultaneously.  We finally compare stochastic vs static policies and to what degree it improves the model.

\medskip
\begin{figure}
    \centering
    \includegraphics[width=90mm,scale=0.5]{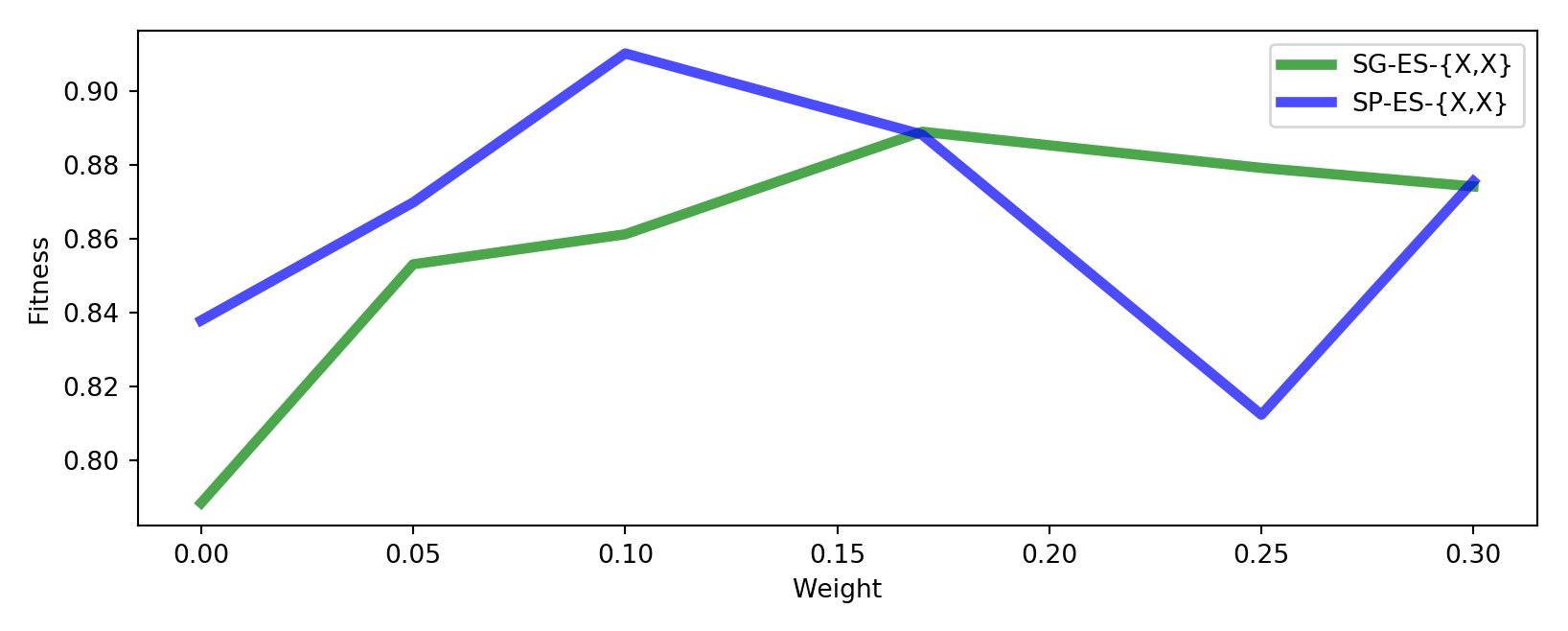}
    \caption{Gini - SG-ES vs. SP-ES}
    \label{fig:gini-gp}
\end{figure}

\noindent
{\bfseries Comparison to Baselines}: Unsurprisingly, we see in Table \ref{Tab:allvar} that while MMR does best with respect to ERR-IA and NDCG, it is at the expense of both Incentives and Gini index which it is unable to optimize.  Similarly, while the Pointwise-ES baseline does well on Gini and Incentives, it performs worse of all the policies on group level diversity, which is not surprising due to document independence assumptions. Importantly, all ES policies were able to optimize all metrics compared to the baseline, providing evidence of its efficacy.

\medskip

\begin{table}[ht]
\scriptsize
\caption{Policy Comparison of Baselines Across All Metrics}
\begin{center}
 \begin{tabular}{l l || c c c } 
 \hline
 Variant (\ref{Tab:weights}) & Metric & Validation & Test Mean & Test Std \\
 \hline
MMR-LambdaMART         & ERR-IA    & 0.487        & 0.480    & - \\
P-ES-\{0.17,0.17\}     & ERR-IA    & 0.481    & 0.467    & - \\
SP-ES-\{0.17,0.17\}    & ERR-IA    & 0.484    & 0.468    & 0.001 \\
G-ES-\{0.17,0.17\}     & ERR-IA    & 0.489    & 0.475    & - \\
SG-ES-\{0.17,0.17\}    & ERR-IA    & 0.478    & 0.471    & 0.002 \\
\hline
MMR-LambdaMART         & Gini      & 0.795    & 0.800    & - \\
P-ES-\{0.17,0.17\}\textbf{*}     & Gini      & 0.925    & 0.891    & - \\
SP-ES-\{0.17,0.17\}\textbf{*}    & Gini      & 0.894    & 0.888    & 0.029 \\
G-ES-\{0.17,0.17\}\textbf{*}     & Gini      & 0.883    & 0.881    & - \\
SG-ES-\{0.17,0.17\}\textbf{*}    & Gini      & 0.911    & 0.889    & 0.011 \\
\hline
MMR-LambdaMART         & Incentive & 0.396        & 0.403    & - \\
P-ES-\{0.17,0.17\}\textbf{*}     & Incentive & 0.466    & 0.525    & - \\
SP-ES-\{0.17,0.17\}\textbf{*}    & Incentive & 0.466    & 0.518    & 0.002 \\
G-ES-\{0.17,0.17\}\textbf{*}     & Incentive & 0.470    & 0.543    & - \\
SG-ES-\{0.17,0.17\}\textbf{*}    & Incentive & 0.466    & 0.543    & 0.009 \\
\hline
MMR-LambdaMART         & NDCG      & 0.692        & 0.679    & - \\
P-ES-\{0.17,0.17\}     & NDCG      & 0.655    & 0.637    & - \\
SP-ES-\{0.17,0.17\}    & NDCG      & 0.652    & 0.634    & 0.001 \\
G-ES-\{0.17,0.17\}     & NDCG      & 0.662    & 0.651    & - \\
SG-ES-\{0.17,0.17\}    & NDCG      & 0.652    & 0.642    & 0.001 \\
 [1ex] 
 \hline
 \multicolumn{4}{c}{\textbf{*} indicates stat. sig. compared to MMR (P < 0.005).} \\
\end{tabular}
\end{center}
\label{Tab:allvar}
\end{table}
\noindent
{\bfseries Influence on Market Indicators}:
We evaluated variants by adjusting the importance weight for the Gini Index and Incentives in \ref{Tab:stoch}. Compared to the baselines, we were able to progressively and smoothly improve both the Gini Index and Incentive indicators with both stochastic and static Greedy variants.  We find that weight does indeed improve the overall equality of the system compared to its unconstrained form \ref{fig:lorenz}.
\begin{table}[ht]
\caption{Gini Index: Stochastic vs. Static (Variant ES-\{0.4,0\})}
 \begin{tabular}{l||c c c} 
 Variant & Validation & Test Mean & Test Std \\ 
 \hline\hline
 G-ES-\{0.4,0\} & 0.906 & 0.798 & 0.000 \\
\hline
 SG-ES-\{0.4,0\}  & 0.922 & 0.903 & 0.008 \\
\end{tabular}
\label{Tab:stoch40}
\end{table}
\begin{figure}
    \centering
    \includegraphics[width=90mm,scale=0.4]{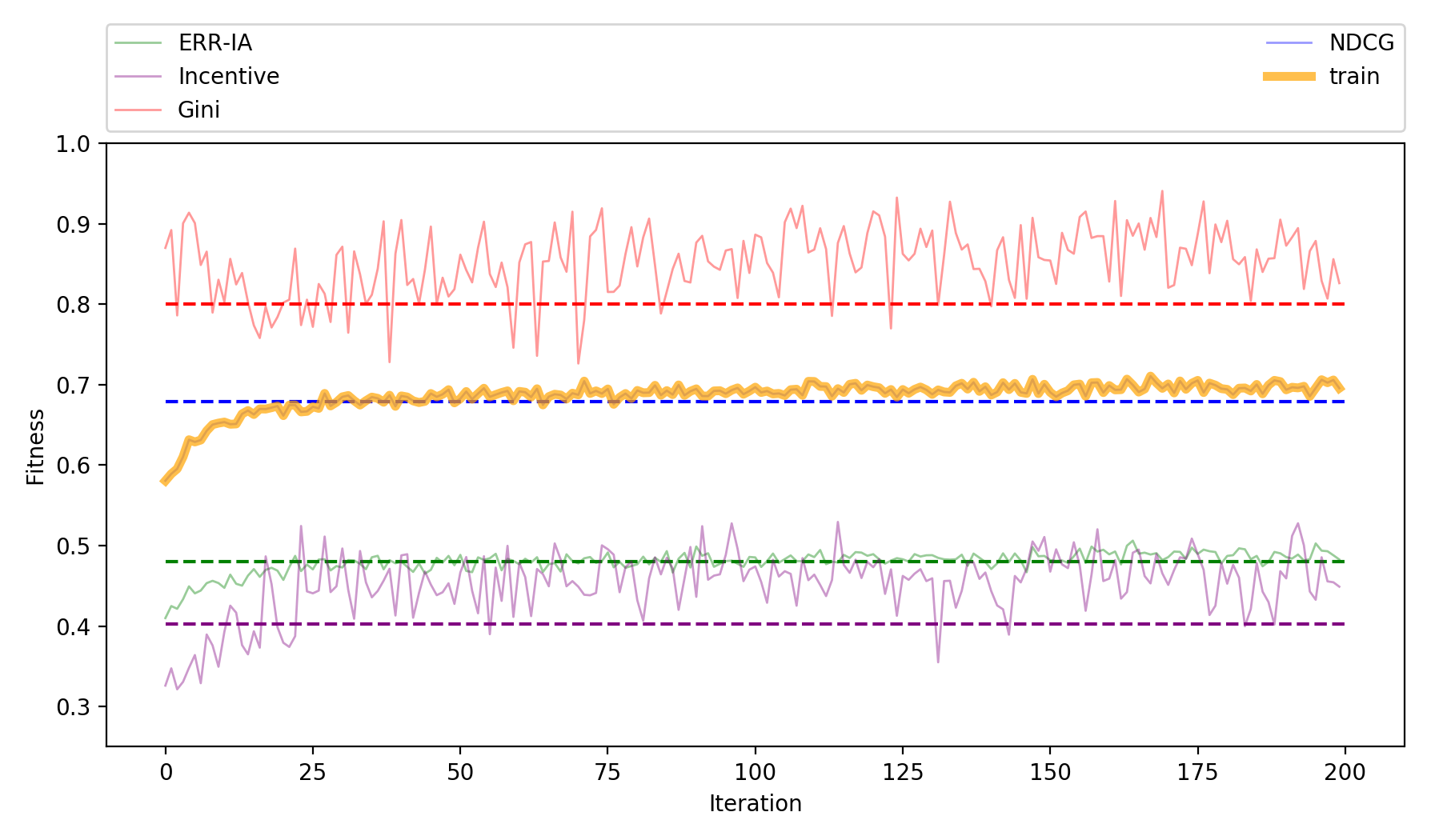}
    \includegraphics[width=90mm,scale=0.4]{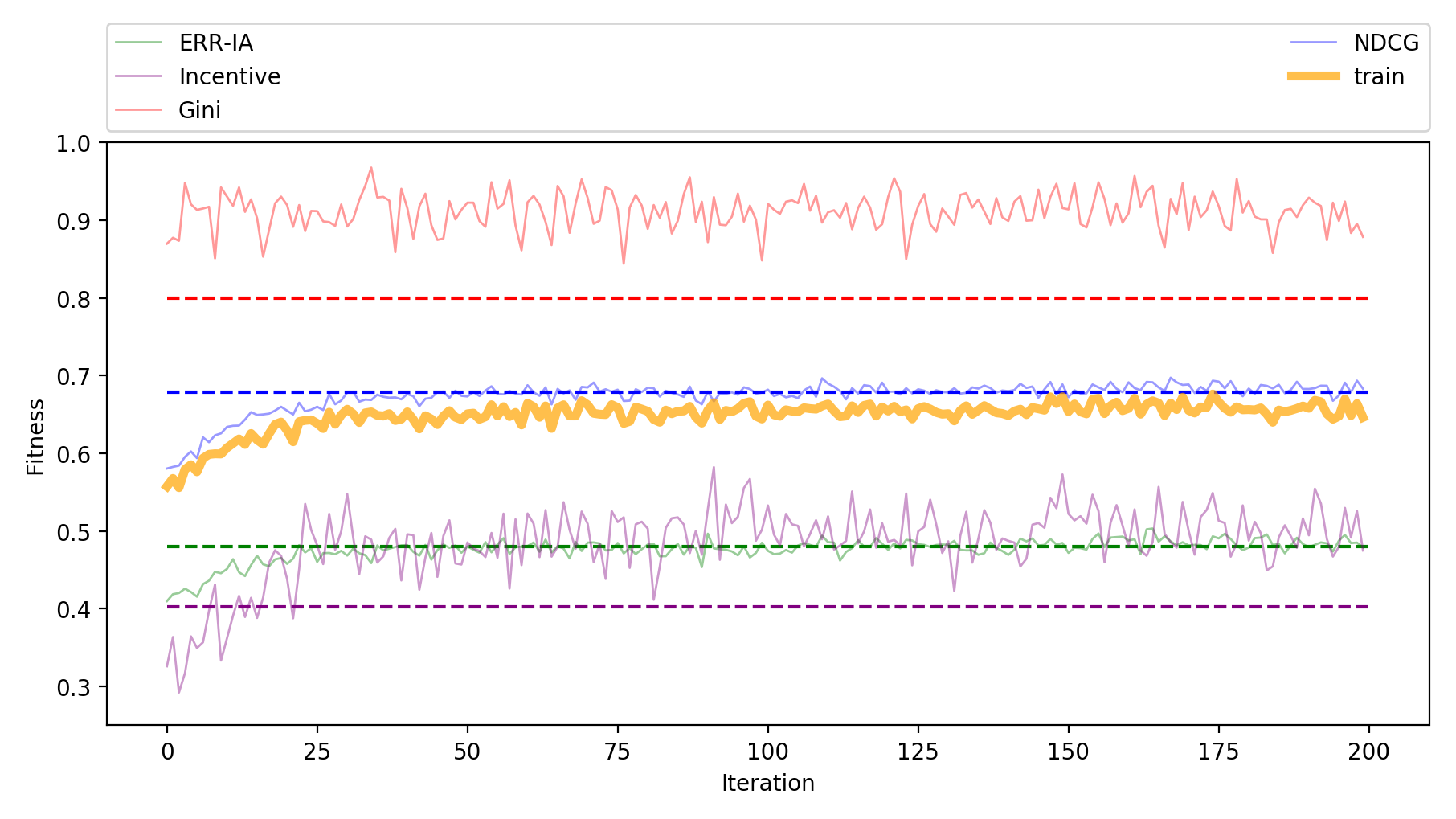}
    \caption{Fitness scores from each iteration - SG-ES-\{0.0,0.0\} (top) and SG-ES-\{0.17,0.17\} (bottom). The horizontal dashed lines display the baseline MMR-LambdaMART scores for each metric matched by color.}
    \label{fig:greedy_iter}
\end{figure}
\medskip

\noindent
{\bfseries Joint Optimization of all Metrics}: We found we were able to optimize multiple metrics simultaneously as seen in Figure \ref{fig:greedy_iter}.  Despite the conflicting nature of the metrics, ES was able to find policies that improved the underlying metrics compared to the baselines.



\medskip

\noindent
{\bfseries Stochastic Features}: Stochastic value functions were competitive with their static brethren.  We find that the additional noise had a few benefits: first, it helped regularize the networks; we find the difference between train and test scores were narrower than the static variants.  Furthermore, the stochastic variants were smoother: they had lower variance as importance weighting increased.  Table \ref{Tab:stoch40} exemplifies the differences when considering only market indicators and relevance.  Table \ref{Tab:stoch} shows how the stochastic policy is more reliable with generalization from validation to the test set. Unlike the greedy variant, SVFs applied to pointwise models were unable to stably improve market indicators.  Figure \ref{fig:gini-gp} compares stochastic pointwise and greedy algorithms on different importance weightings - while greedy is fairly smooth, the pointwise model displays high variance both within policy and across different weightings.
\begin{table}[ht]
\caption{Gini Index: Stochastic vs. Static}
\begin{center}
 \begin{tabular}{l|| c c c} 
 Variant (\ref{Tab:weights}) & Validation & Test Mean & Difference \\ 
 \hline
 G-ES-\{0,0\} & 0.740 & 0.849  & -0.109 \\
SG-ES-\{0,0\}  & 0.734 & 0.789 & -0.065 \\
\hline
 G-ES-\{0,0.05\} & 0.815 & 0.883 & -0.068 \\
SG-ES-\{0,0.05\}  & 0.854 & 0.853 & 0.001 \\
\hline
 G-ES-\{0,0.1\} & 0.825 & 0.899 & -0.074 \\
SG-ES-\{0,0.1\} & 0.881 & 0.861 & 0.020 \\
\hline
 G-ES-\{0,0.17\} & 0.883 & 0.881 & 0.002 \\
SG-ES-\{0,0.17\} & 0.911 & 0.889 & 0.012 \\
\hline
 G-ES-\{0,0.25\} & 0.924 & 0.861 & 0.063 \\
SG-ES-\{0,0.25\} & 0.901 & 0.879 & 0.022 \\
\hline
 G-ES-\{0,0.3\} & 0.945 & 0.876 & 0.069 \\
SG-ES-\{0,0.3\} & 0.925 & 0.874 & 0.051 \\
\end{tabular}
\end{center}
\label{Tab:stoch}
\end{table}

\section{Conclusion and Future Work}
In this paper we defined types of market indicators critical for creating healthy, two-sided marketplaces and proposed strategies for learning policies to jointly maximize those desired market characteristics. We showed that we can influence these market-level metrics via our models, resulting in a method for imposing business needs while eliminating many of the common forms of interventions that lead to sub-par search experiences.

There are many possible directions for future work. One is to explore stochastic models in online environments versus just offline demonstrated in this paper. Another possible direction is to explore more efficient algorithms as ES comes at significant cost of sample efficiency, often using orders of magnitude more compute than other, more efficient policy gradient optimizers. 
The last one is to see how these ideas can be utilized in a production system.



%
\bibliographystyle{ACM-Reference-Format}
\bibliography{sample-base}

\end{document}